\documentclass{article}

\usepackage{graphicx}
\usepackage{xcolor}
\usepackage[version=3]{mhchem}
\usepackage{epstopdf}
\usepackage{amssymb}

\usepackage{amsthm}
\usepackage{thmtools}
\usepackage{mdframed}

\theoremstyle{definition}
\newmdtheoremenv{exercise}{Exercise}

\providecommand{\detail}[1]{#1}
\renewcommand{\detail}[1]{}

\begin{document}

\title{A toy model solved by statistical
  mechanics for teaching reaction kinetics beyond ideality.}

\author{Doriano Brogioli, Fabio La Mantia}
\date{Universit\"at Bremen, Energiespeicher-- und
  Energiewandlersysteme, Bibliothekstra{\ss}e 1, 28359 Bremen,
  Germany. E-mail: brogioli@uni-bremen.de}

\maketitle

\begin{abstract}
  Chemical equilibrium is fully characterized by thermodynamics, while
  the rates of chemical reactions can be calculated for ideal
  solutions by using mass-action equations. The evaluation of the
  rates of reactions in a non-ideal system is instead much more
  complex, even at equilibrium, being dependent on the microscopic
  features of the interactions: no universal theory exist.  Here we
  propose a toy model to help students understand such complexity. It
  is formulated by means of statistical mechanics and aims at the
  evaluation of the exchange reaction rate at equilibrium. The toy
  model can be solved by the students, with various types of
  interactions, as an exercise. The results prove that no general rule
  connects the reaction rates to the thermodynamic quantities, such as
  the activity coefficients, dramatically proving the complexity and
  richness of the field.
\end{abstract}


\section{Introduction}

In the chemistry courses, the kinetics of chemical reactions is mostly
discussed for \emph{ideal} solutions~\cite{atkins}, for which the rate
$r$ of the reaction
\begin{equation}
  n_A \mathrm{A} + n_B \mathrm{B} + \dots  \to \dots
\end{equation}
is characterized by the mass-action equations:
\begin{equation}
  r = k c_A^{n_A} c_B^{n_B} \dots
  \label{eq:intro:mass:action}
\end{equation}
where the $c_j$ are the concentrations of the reactants, $n_j$ are the
stoichiometric coefficients of the reaction, and $k$ is the reaction
rate constant. This kind of equations is based on
several assumptions, which are however valid in significant cases.

The possible non-ideal behaviour of the system, i.e. the presence of
interactions between particles, is instead easily handled by
thermodynamics, thus at chemical \emph{equilibrium}, e.g.
\begin{equation}
  n_A \mathrm{A} + n_B \mathrm{B} + \dots  \leftrightharpoons
  n_C \mathrm{C} + n_D \mathrm{D} + \dots 
\end{equation}
For such a reaction, the equilibrium is expressed in terms
of the chemical potentials $\mu_j$~\cite{atkins}:
\begin{equation}
  n_A \mu_A + n_B \mu_B + \dots = n_C \mu_C + n_D \mu_D + \dots
  \label{eq:intro:thermo:equilibrium}
\end{equation}

\begin{figure}
  \begin{center}
    \includegraphics{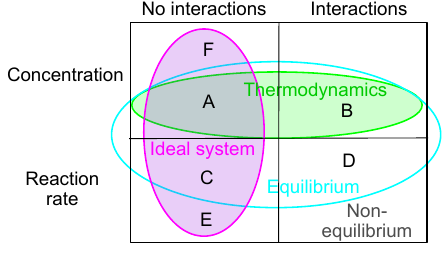}
  \end{center}
  \caption{Classification of theories describing reaction equilibrium
    and kinetics. }
  \label{fai:overview}
\end{figure}

Figure~\ref{fai:overview} schematically shows these concepts.  The
theory of ideal systems deals with systems in which there is no
particle interaction. It enables the calculation of the reaction rates
and time-dependence of the concentrations, even outside equilibrium
(regions E and F, respectively). Under equilibrium, where the net
reaction rate vanishes, the theory predicts the exchange reaction
rates (region C) together with the equilibrium concentrations (region
A). Thermodynamics holds at equilibrium and enables the calculation of
the concentrations of chemical species, even in the presence of
interactions (regions A and B).

The two calculation schemes overlap nicely within their shared
validity range (region A), i.e. for systems that are both at
equilibrium and ideal, when dealing with concentrations. In ideal
solutions, the chemical potentials are:
\begin{equation}
  \mu_j=\mu_j^0+k_BT \ln \frac{c_j}{c_0} \, ,
  \label{eq:approx:mu:ideal:c}
\end{equation}
where $k_B$ is the Boltzmann constant, $T$ is the temperature,
$\mu^0_j$ is the standard chemical potential of the $j$th species, and
$c_0$ is the standard concentration. Using this expression, the
thermodynamic equilibrium condition
(Eq.~\ref{eq:intro:thermo:equilibrium}) and the mass-action equations
(Eq.~\ref{eq:intro:mass:action}) merge to give:
\begin{equation}
  R = k_f c_A^{n_A} c_B^{n_B} \dots = k_b c_C^{n_C} c_D^{n_D} \dots
  \label{eq:intro:eq:ideal:kf:fb}
\end{equation}
where $R$ is the exchange reaction rate, equal to both the forward and
backward reaction rates; $k_f$ and $k_b$ are the forward and backward
reaction rate constants, respectively, related by:
\begin{equation}
  \frac{k_f}{k_b} = e^{-\frac{\Delta G^0}{k_BT}}
  \label{eq:intro:ratio:kf:kb}
\end{equation}
where 
\begin{equation}
  \Delta G^0 =
  n_C \mu_C^0+n_D \mu_D^0+\dots
  -\left(
  n_A \mu_A^0+n_B \mu_B^0+\dots
  \right)
  \label{eq:intro:Delta:G:0}
\end{equation}
is the standard Gibbs free energy of reaction.

We now focus on region D of Fig.~\ref{fai:overview}, where chemical
equilibrium holds but the system is not ideal, i.e. the interactions
are not negligible. Under these conditions, the thermodynamic
quantities, including concentrations and their fluctuations, can be
calculated, but there is no general theory for calculating the
reaction kinetics (exchange rate). The reason is that thermodynamics
has the concept of ``sequence'' but no concept of time.  A fundamental
tool in statistical mechanics, also valid in region D, is the
fluctuation-dissipation theorem~\cite{lebellac}, which will be shortly
discussed below. Rigorous treatments for the evaluation of rates are
only valid in specific cases. This lack of a general scheme represents
an obstacle not only in research, but also in teaching: the message
that must be conveyed is that care must be taken to use the right tool
for each case and to avoid misleading generalizations and
simplifications. A good understanding of the region D is also a useful
training before facing the so-called non-equilibrium thermodynamics,
outside the regions A, B, C and D, which presents additional
challenges and is outside the scope of this paper.

The first problem that we encounter in extending
Eq.~\ref{eq:intro:eq:ideal:kf:fb} outside ideality is that the second
equality fails. Indeed, we know that the correct relation is:
\begin{equation}
  k_f a_A^{n_A} a_B^{n_B} \dots = k_b a_C^{n_C} a_D^{n_D} \dots
  \label{eq:intro:equilibrium:activities:kf:fb}
\end{equation}
where $a_j$ is the activity of the $j$th species. This equation could thus
suggest to extrapolate Eq.~\ref{eq:intro:eq:ideal:kf:fb} to the
questionable (and erroneous in general) equation:
\begin{equation}
  R \stackrel{?}{=} k_f a_A^{n_A} a_B^{n_B} \dots = k_b a_C^{n_C} a_D^{n_D} \dots
  \label{eq:intro:erroneous}
\end{equation}
Undoubtedly, there are two good points in favour of this equation: i)
it approaches the ideal equation for vanishing concentrations; ii) the
net reaction rate (the difference between second and third term)
vanishes at equilibrium. However, the first equality of
Eq.~\ref{eq:intro:erroneous} is only an extrapolation, which is not
correct in general. The main question discussed in this paper is if
Eq.~\ref{eq:intro:erroneous} provides an evaluation of the
exchange rate $R$ that is better than
Eq.~\ref{eq:intro:eq:ideal:kf:fb}, when applied outside idelity; in
general, the answer is negative.

Education in chemistry should prevent the above-described unjustified
extrapolations.  The courses should clarify that the correct
evaluation of the effect of interactions (non-ideality) on the
reaction rates requires a suitable approach, lying outside the realm
of thermodynamics: chemical potentials, activities, and activity
coefficients are not enough to characterize the chemical kinetics. The
substitution of concentrations, like in
Eq.~\ref{eq:intro:eq:ideal:kf:fb}, with activities, like in
Eq.~\ref{eq:intro:erroneous}, does not lead to a better evaluation of
the effects of interactions; taken alone, this substitution can lead
even more far from the correct result.

A successful approach for predicting reaction rates is the transition
state theory~\cite{truhlar1996, perezbenito2017}, which models the
passage of a barrier. In this theory, the Gibbs free energy of the states
and of the so-called transition state are taken into consideration,
and thus activities play a role in determining the \emph{net} reaction
rate~\cite{baird1999}, at least, under the hypothesis needed for the
application of this theory. This notwithstanding, this approach does
not justify the evaluation of the \emph{exchange} reaction rate
Eq.~\ref{eq:intro:erroneous} outside ideality.

Fundamental results, valid at equilibrium, have been obtained by means
of the fluctuation-dissipation theorem~\cite{lebellac}.  In general,
the theorem relates fluctuations, which are thermodynamic quantities,
to kinetic properties. In the case of chemical
reactions\cite{baird2003}, it relates the exchange rate and the
concentration fluctuations to the decay rate of concentration
fluctuations; the latter rate can be used to roughly evaluate the net
reaction rate around equilibrium. Notwithstanding the importance of
such results, the theorem does not enable the independent
determination of the two rates; a microscopic description of the
system remains necessary.

In this paper, we develop a toy model that represents a reaction at
equilibrium in the presence of interactions, described in terms of
statistical mechanics. We propose it as an educational tool for
undergraduate students in chemistry with a minimal knowledge of
statistical mechanics.

We derive general expressions of the exchange reaction rate and of the
chemical potentials in terms of averages on the ensemble distribution.
Then, we apply them to various forms of interaction among
particles. The applications of the general formula to specific cases
can be carried out, as an exercise, by the students. The prerequisite
is a minimal knowledge of statistical mechanics; students with a
background in numerical calculation can also easily perform Monte
Carlo simulations.

The aim is to show that both the chemical potentials (and thus the
activity coefficients) and the reaction rate depend on the
interactions, but the mathematical expression of this dependence
strongly changes among the various examples. The results dramatically
prove that the thermodynamic parameters (chemical potentials or
activity coefficients) are not able to characterize the kinetics in
general, not even close to equilibrium.

\section{Description of the toy models and general solution}

\subsection{Description of the system and notations}
\label{sect:notation}

\begin{figure}
  \begin{center}
    \begin{tabular}{cc}
      \parbox[b][5cm][c]{5mm}{(a)} &
      \includegraphics{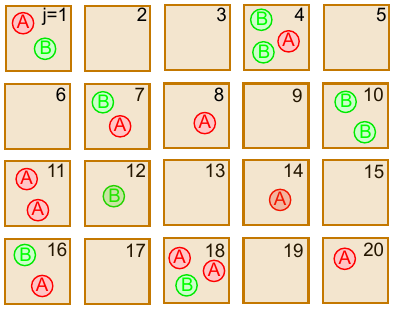} \\
      \parbox[b][5cm][c]{5mm}{(b)} &
      \includegraphics{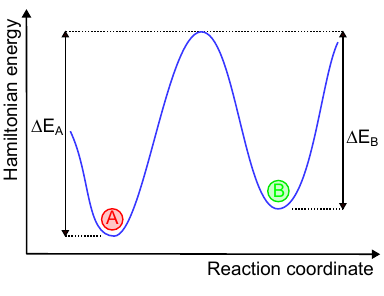}
    \end{tabular}
  \end{center}
  \caption{Sketch of the model system. Panel (a): each cell can
    contain any number of particles, indexed by $j$. Panel (b):
    value of Hamiltonian (dotted) and discrete states A
    and B, representing the different chemical species. }
  \label{fig:geometry}
\end{figure}

We consider an equilibrium between two species A and B:
\begin{equation}
  \mathrm{A} \rightleftharpoons \mathrm{B}
  \label{eq:reaction}
\end{equation}
Figure~\ref{fig:geometry}a shows a sketch of the system. The two
species A and B are in a lattice with $N$ cells; in general there can
be any number of particles in each cell. We will focus on interactions
among particles in the same cell, thus the physical, geometrical
arrangement of the cells (e.g. cubic lattice) does not play a role. A
single index $j$ runs along all the cells of the lattice.

A configuration of the system is represented by an entity $s$
consisting in two arrays, $s^T$ and $s^P$: $s^T_i$ is either A or B
and represents the type of the $i$th particle, and $s^P_i$ is the
position of the $i$th particle in the lattice, i.e. a natural number
between 1 and $N$. We denote $\left|s\right|$ the number of elements
of the arrays; $N_A(s)$ and $N_B(s)$ are the number of particles A and
B, respectively, in the state $s$.

The calculations will be done in a canonical setup, with a fixed $N_T$
number of particles. Hence, the possible states of the system are
represented by states $s$ such that $|s|=N_T$. It must be remarked
that, during the calculations, states with different length $|s|$ will
appear.

The notation $s+\left<A,j\right>$ represents a state $s$ to which an
additional particle A has been added at position $j$. An analogous
notation is used for adding a particle B. It must be noticed that the
number of particles increases by 1,
i.e. $|s+\left<A,j\right>|=|s|+1$. The notation $s|s^T_i=A$ means that
the $i$th particle is substituted by a particle A, while keeping its
position in cell $s_i^P$. In this latter case, the number of particles
does not change, i.e. $|(s|s^T_i=A)|=|s|$.

The system is specified by giving the Hamiltonian function
$H\left(s\right)$. This Hamiltonian treats A and B as discrete states
and can be seen as an approximation of a continuous model, smoothly
interpolating between the two states A and B along a reaction
coordinate. As an example,
Fig.~\ref{fig:geometry}b shows the dependence of the Hamiltonian
energy on the reaction coordinate, with a barrier between A and B.

According to Boltzmann equation, the average on the canonical ensemble
of $N_T$ particles of a variable $a$ is:
\begin{equation}
  \left<a\right>_{N_T} = \frac{1}{Z_{N_T}}
  \sum_{s, \left|s\right|=N_T} a\left(s\right)
  e^{-\frac{H\left(s\right)}{k_BT}}
  \label{eq:def:canonical:average}
\end{equation}
where $k$ is the Boltzmann constant and $T$ is the temperature of the
system, and $Z_{N_T}$ is the partition function:
\begin{equation}
  Z_{N_T} = \sum_{s, \left|s\right|=N_T}
  e^{-\frac{H\left(s\right)}{k_BT}}
  \label{eq:def:Z}
\end{equation}

\subsection{Definition of reaction rate}

The reaction that we are taking into account has a single step and
does not pass through intermediate states, such as activated chemical
species. This assumption is not a limitation; rather, it means that we
focus on a single step of a reaction, while the overall process could
involve multiple steps and intermediate (possibly transition)
states.

We model the reaction as the passage of a barrier.  Our approach is
based on the Arrhenius law~\cite{arrhenius1889, arrhenius1889bis,
  laidler1984}, which provides an empirical evaluation of the
dependence of the reaction constant on the temperature:
\begin{equation}
  k = \Lambda e^{-\frac{\Delta E}{k_BT}} ,
\end{equation}
where $\Lambda$ and the energy barrier $\Delta E$ are two empirical
parameters.

Two different approaches have been proposed in order to justify the
Arrhenius law: i) in the collision theory~\cite{divincenzo2020},
$\Delta E$ is interpreted as a mechanical potential energy barrier;
ii) in the transition state theory~\cite{truhlar1996} (or activated
state theory), $\Delta E$ is interpreted as a free energy barrier,
separating the reactants from an activated state. The two approaches
are rigorously incompatible with each other, since they attribute a
different nature to the same empirical parameter $\Delta E$.

In our model, the barrier is described by a Hamiltonian, representing
the mechanical energy.  The forward and backward reaction rates are
expressed as:
\begin{align}
  r_{A\to B} & = \left< \sum_i^{N_T} \delta_{s_i^T=A} \Lambda
  \frac{N_A}{N} e^{-\frac{\Delta E_A(s,i) }{k_BT}} \right>_{N_T}
  \label{eq:arrhenius:rate:A:B} \\
  r_{B\to A} & = \left< \sum_i^{N_T} \delta_{s_i^T=B} \Lambda
  \frac{N_B}{N} e^{-\frac{\Delta E_B(s,i) }{k_BT}} \right>_{N_T}
  \label{eq:arrhenius:rate:B:A}
\end{align}
where where $\delta_{\mathrm{condition}}$ is 1 if ``condition'' is
met, 0 otherwise, $\Delta E_A(s,i)$ and $\Delta E_B(s,i)$ are the
heights of the energy barrier from the side of the state A and B,
respectively, in state $s$, for the $i$th particle (see
Fig.~\ref{fig:geometry}), and $\Lambda$ is a linear dependence factor
(see Sect.~\ref{sect:notation} for the notation $s|s^T_i=A$).

The interactions affect the energy of the particles, thus also the
height of the energy barriers, $\Delta E_A(s,i)$ and $\Delta E_B(s,i)$. We can
calculate the difference $\Delta E_A(s,i)-\Delta E_B(s,i)$ from statistical
mechanics, since it equals the difference of energy between states A
and B:
\begin{equation}
  \Delta E_A(s,i) - \Delta E_B(s,i)  =
  H\left(s|s^T_i=B\right)-H\left(s|s^T_i=A\right)
\end{equation}
In order to separately calculate $\Delta E_A(s,i)$ and $\Delta
E_B(s,i)$, we assume that the known energy variation
$H\left(s|s^T_i=B\right)-H\left(s|s^T_i=A\right)$ is linearly
distributed among the two particles A and B:
\begin{align}
  \Delta E_A & = \Delta E+\left(1-\alpha\right) 
  \left[ H\left(s|s^T_i=B\right)-H\left(s|s^T_i=A\right) \right]
  \label{eq:arrhenius:single:particle:A}
  \\
  \Delta E_B & = \Delta E+\alpha 
  \left[ H\left(s|s^T_i=A\right)-H\left(s|s^T_i=B\right) \right]
  \label{eq:arrhenius:single:particle:B}
\end{align}
where $\Delta E$ is a fixed energy barrier and $\alpha$ is a linear
distribution factor, $0\le \alpha\le 1$. This linear distribution is
actually observed in transition state theory~\cite{atkins}, in the
form of a linear dependence of the height of the barrier on $\Delta
G^0$, known as the Evans-Polanyi principle. A similar linear
dependence also appears in the derivation of the Butler-Volmer
equation of electrochemistry~\cite{bard}, where the coefficient
$\alpha$ is called the charge transfer coefficient.  The linear
splitting can also be justified under the approximation of smooth
dependence of the Hamiltonian on the reaction coordinate.

Eqs.~\ref{eq:arrhenius:rate:A:B} and
\ref{eq:arrhenius:rate:B:A} are then rewritten as:
\begin{align} r_{A\to B} & = \label{eq:rate:A:B:def}
  \left< \sum_{i=1}^{N_T}
  \delta_{s_i^T=A}
  \frac{\Lambda}{N} e^{-\frac{\Delta E+\left(1-\alpha\right)
      \left[H\left(s|s^T_i=B\right)-H\left(s\right)\right]}{k_BT}}
  \right>_{N_T}
  \\  r_{B\to A} & = \label{eq:rate:B:A:def}
  \left< \sum_{i=1}^{N_T}
  \delta_{s_i^T=B}
  \frac{\Lambda}{N} e^{-\frac{\Delta E+\alpha
      \left[H\left(s|s^T_i=A\right)-H\left(s\right)\right]}{k_BT}}
  \right>_{N_T}
\end{align}
Although this model of reaction rate is quite simple, it is
enough to discuss a wide range of phenomena.

\subsection{Expression of reaction rates and activity coefficients}

The reaction rates and the activity coefficients will be expressed in
terms of two quantities, $\Gamma_A(s,j)$ and $\Gamma_B(s,j)$,
representing the interactions in the $j$th cell when the system is in
state $s$ :
\begin{align}
  \Gamma_A\left(s,j\right) & = 
  e^{-\frac{H\left(s+\left<A,j\right>\right)-H\left(s\right)}{k_BT}}
  \label{eq:def:GammaA:main} \\
  \Gamma_B\left(s,j\right) & = 
  e^{-\frac{H\left(s+\left<B,j\right>\right)-H\left(s\right)}{k_BT}}
  \label{eq:def:GammaB:main}
\end{align}
(see Sect.~\ref{sect:notation} for the notation $s+\left<A,j\right>)$
Intuitively, the two factors represent the variation of probability of
the state $s$ when a particle A or B is inserted in cell $j$. We
assume that the result is actually independent of the cell position,
so that $j$ can be taken arbitrarily.

We write the chemical potentials $\mu_A$ and $\mu_B$ as:
\begin{align}
  \mu_A & = \mu_A^0 + k T \ln \left( \gamma_A \frac{\bar{N}_A}{N} \right)
  \label{eq:def:mu:A:lambda:main}
  \\
  \mu_B & = \mu_B^0 + k T \ln \left( \gamma_B \frac{\bar{N}_B}{N} \right)
  \label{eq:def:mu:B:lambda:main}
\end{align}
where $\bar{N}_A$ and $\bar{N}_B$ are the average numbers of particles
(the actual numbers of particles fluctuate from sample to sample),
$\mu_A^0$ and $\mu_B^0$ are the standard chemical potentials, and
$\gamma_A$ and $\gamma_B$ are the activity coefficients, representing
the deviation of the activities from ideality.  In Supporting
Information, SI-Sect.~\ref{appendix:activity:coefficients}, we give an
expression of the standard chemical potentials and of the
activity coefficients in terms of averages of the functions $\Gamma_A$
and $\Gamma_B$:
\begin{align}
  \mu_A^0 & = -k_BT \ln \left< \Gamma_A\left(s,j\right) \right>_{0}^j
  \label{eq:def:mu:A:zero:main}
  \\
  \mu_B^0 & = -k_BT \ln \left< \Gamma_B\left(s,j\right) \right>_{0}^j
  \label{eq:def:mu:B:zero:main}
  \\
  \gamma_A & = \frac
        { \left< \Gamma_A\left(s,j\right) \right>_{0}^j }
        { \left< \Gamma_A\left(s,j\right) \right>_{N_T}^j }
  \label{eq:def:gamma:A:main}
  \\
  \gamma_B & = \frac
        { \left< \Gamma_B\left(s,j\right) \right>_{0}^j }
        { \left< \Gamma_B\left(s,j\right) \right>_{N_T}^j }
  \label{eq:def:gamma:B:main}
\end{align}
In all the proposed exercises, $\mu_A^0=\mu_B^0=0$; in the following, we
assume that this simplifying assumption holds, also
implying
$\left< \Gamma_A\left(s,j\right) \right>_{0}^j = \left< \Gamma_B\left(s,j\right) \right>_{0}^j=1$.

In SI-Sect.~\ref{appendix:elaboration:reaction:rate}, we calculate the
so-called exchange reaction rate $R=r_{A\to B}=r_{B\to A}$; the
forward and backward reaction rates are indeed equal, as it can be
shown from Eqs.~\ref{eq:rate:A:B:def} and \ref{eq:rate:B:A:def} and
expected in thermodynamic equilibrium. The exchange reaction rate is:
\begin{equation}
  R = \frac{\Lambda}{2} e^{-\frac{\Delta E}{k_BT}} \vartheta \frac{N_T}{N}
  \label{eq:R:xi:main}
\end{equation}
where
\begin{equation}
  \vartheta = 2
  \frac{
    \left< \Gamma_A\left(s,j\right)^{\alpha} \Gamma_B\left(s,j\right)^{1-\alpha} \right>_{N_T}^j
  }{
    \left< \Gamma_A\left(s,j\right) \right>_{N_T}^j
    +
    \left< \Gamma_B\left(s,j\right) \right>_{N_T}^j
  }
  \label{eq:xi:main}
\end{equation}
We remark once again that these expressions hold for
$\mu_A^0=\mu_B^0=0$, like in the exercises below.

\subsection{Meaning of the parameter $\vartheta$}

From Eq.~\ref{eq:xi:main} we notice that $\vartheta=1$ for an ideal
system. This suggests to interpret $\vartheta$ as a factor representing the
deviation of the reaction rate from ideality.

For reaction Eq.~\ref{eq:reaction}, the equilibrium condition is:
\begin{equation}
  e^{\frac{\mu_A^0}{k_BT}} \gamma_A c_A =
  e^{\frac{\mu_B^0}{k_BT}} \gamma_B c_B  
\end{equation}
This equation enables the calculation of the concentrations $c_A$ and
$c_B$ as a function of the total concentration $c_T$. In turn, these
concentrations can be used to try to evaluate $R$ through
Eq.~\ref{eq:intro:erroneous}:
\begin{equation}
  R \stackrel{?}{=} k
  \frac
      {\left( e^{\frac{\mu_A^0}{k_BT}} + e^{\frac{\mu_B^0}{k_BT}} \right)
        \gamma_A \gamma_B}
      {e^{\frac{\mu_A^0}{k_BT}} \gamma_A + e^{\frac{\mu_B^0}{k_BT}} \gamma_B}
      c_T
  \label{eq:erroneous:R:k}
\end{equation}
where
\begin{equation}
  k = k_f \frac
  {e^{\frac{\mu_B^0}{k_BT}}}
  {e^{\frac{\mu_A^0}{k_BT}} + e^{\frac{\mu_B^0}{k_BT}}}
  = k_b \frac
  {e^{\frac{\mu_A^0}{k_BT}}}
  {e^{\frac{\mu_A^0}{k_BT}} + e^{\frac{\mu_B^0}{k_BT}}}
\end{equation}
is a single constant defining the exchange reaction rate. The
comparison of Eq.~\ref{eq:erroneous:R:k} to Eq.~\ref{eq:R:xi:main}
suggests to write $\vartheta$ as:
\begin{equation}
  \vartheta \stackrel{?}{=}
  \frac
      {2 \gamma_A \gamma_B}
      {\gamma_A + \gamma_B}
  \label{eq:erroneous:check}
\end{equation}
under the assumption $\mu_A^0=\mu_B^0=0$; it is the harmonic mean of
the activity coefficients. Being a consequence of
Eq.~\ref{eq:intro:erroneous}, evaluating the validity of this equation
corresponds to investigate the validity of the use of activity
coefficients to calculate reaction rates. The equation in question,
Eq.~\ref{eq:erroneous:check}, depends on the non-ideal effects through
the activities, according to Eq.~\ref{eq:erroneous:R:k} and, in turn,
to Eq.~\ref{eq:intro:erroneous}; the question is if it is a better
approximation of $\vartheta$ than simply taking the value in ideal
systems, $\vartheta=1$. The answer will be that it is not, not even
qualitatively.

By comparing Eq.~\ref{eq:xi:main} with Eqs.~\ref{eq:def:gamma:A:main}
and \ref{eq:def:gamma:B:main}, it can be noticed that $\vartheta$ and the
$\gamma_j$ are expressed in terms of averages of functions of the
$\Gamma_j$. This concept expresses the idea that the reaction rate and
the thermodynamic variables are altered by the interactions.  However,
the averages and the functions are different, thus the expressions
suggest that it is impossible to derive $\vartheta$ from the $\gamma_j$.

From the definition of $\vartheta$, Eq.~\ref{eq:xi:main}, in the case
$\alpha=1/2$ (as in the exercises below), we find that $0\le \vartheta \le
1$. The lower bound is trivial; the upper bound is proved in
SI-Sect.\ref{si:sect:bound:xi}.  The upper bound is tight, since it is
reached by ideal solutions, i.e. in the absence of interactions
(constant $H$). From these bounds, it is immediately clear that
Eq.~\ref{eq:erroneous:check} is not true in general: the right-hand
side can be larger than 1, while $\vartheta$ is smaller than 1.

\section{Exercises}

In this section, we describe some possible exercises. In each, a
specific form of interaction is described, defining a Hamiltonian. The
exercise consist in:
\begin{itemize}
\item calculating the parameters $\gamma_j$ and $\vartheta$;
\item qualitatively explaining the trends;
\item checking if the equation in question,
  Eq.~\ref{eq:erroneous:check}, holds.
\end{itemize}

The calculation can be done symbolically or by Monte Carlo numerical
methods, depending on the knowledge of the students. In all the cases
it can be found that $\mu_A^0=\mu_B^0=0$; moreover the value
$\alpha=1/2$ is assumed.

\begin{figure}
\includegraphics{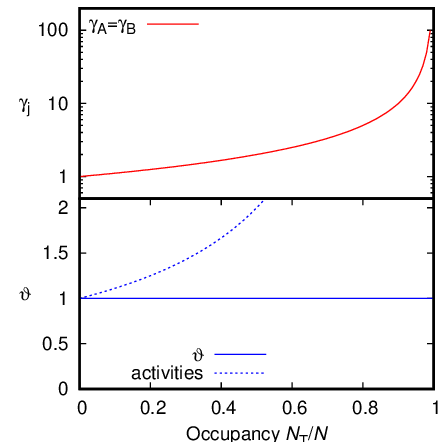}
  \caption{Results of Exercise.~\ref{sect:ex:excluded:volume:all}, on
    excluded volume interaction among all the particles. The dashed
    curve, labelled as ``activities'', is the expectation from the
    equation in question, Eq.~\ref{eq:erroneous:check}. }
  \label{fig:ex:exluded:volume:all}
\end{figure}

\begin{exercise}[Excluded volume interaction among all the particles]
\label{sect:ex:excluded:volume:all}
We assume that any particle excludes the presence
of other particles in the same cell (whatever is the species).  No
other interaction is present. We model this behaviour by assuming that
the Hamiltonian takes a value 0 on the allowed configurations and a
very large value otherwise, such that $e^{-H}$ vanishes.
\end{exercise}

\begin{proof} We can immediately see that
$\Gamma_A\left(s,j\right)=\Gamma_B\left(s,j\right)$ is 1 on allowed
configurations $s$ if the cell $j$ is empty in configuration $s$, and
it is 0 otherwise. It is easy to find the following solutions:
\begin{align}
  \gamma_A & = \frac{N}{N-N_T} \\
  \gamma_B & = \frac{N}{N-N_T} \\
  \vartheta & = 1
\end{align}
The results are shown in Fig.~\ref{fig:ex:exluded:volume:all}. The
increase of the activity coefficient with the number of particles is
related to the repulsivity of the force. The reaction rate is not
affected by the interaction: the transition of a particle from A to B
or \emph{vice-versa} does not change the interaction energy. The
failure of Eq.~\ref{eq:erroneous:check} is dramatic, since $\vartheta$
remains constant as a function of $N_T$, keeping the ideal value,
while the equation in question predicts an increase:
\begin{equation}
  \frac{2 \gamma_A \gamma_B}{\gamma_A + \gamma_B} = \frac{N}{N-N_T}
\end{equation}
We thus see that, in this case, the correct value of $\vartheta$
equals the ideal one, while the use of activity coefficients,
according to Eq.~\ref{eq:erroneous:check}, brings the result to a
wrong value.
\end{proof}

\begin{figure}
\includegraphics{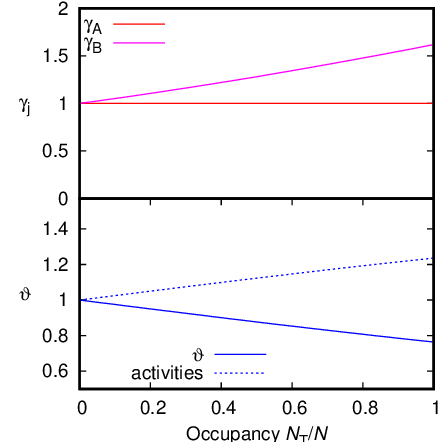}
  \caption{Results of Exercise~\ref{sect:ex:excluded:volume:single},
    on excluded volume interaction among particles of species B. The dashed
    curve, labelled as ``activities'', is the expectation from the
    equation in question, Eq.~\ref{eq:erroneous:check}. }
  \label{fig:ex:exluded:volume:single}
\end{figure}

\begin{exercise}[Excluded volume interaction among particles of a single species]
\label{sect:ex:excluded:volume:single}
At variance with the previous exercise, a particle $B$ excludes other
particles $B$, but any number of particles $A$ can be present in the
same cell, possibly together with a single particle B.
\end{exercise}

\begin{proof}
  For every allowed configurations $s$, $\Gamma_A\left(s,j\right)$ is
  1. Instead, $\Gamma_B\left(s,j\right)$ is 1 if, in configuration
  $s$, the cell $j$ does not contain a $B$, and it is 0
  otherwise. Then we find:
\begin{align}
  \gamma_A & = 1 \\
  \gamma_B & = \frac{N}{N-\bar{N}_B} \\
  \vartheta & = \frac{N-\bar{N}_B}{N-\frac{\bar{N}_B}{2}}
\end{align}

To find the relation between $\bar{N}_B$ and $N_T$, we impose the
equilibrium (see Eqs.~\ref{eq:def:mu:A:lambda:main} and
\ref{eq:def:mu:B:lambda:main}): $N_T=(1+\gamma_B)\bar{N}_B$. The
results are shown in Fig.~\ref{fig:ex:exluded:volume:single};
$\vartheta$, $\gamma_A$, and $\gamma_B$ are plotted with horizontal
coordinate $N_T/N$, parametrically in $\bar{N}_B$.

In this case, $\vartheta$ is decreased by the interaction: in a cell
with one particle A and one particle B, the transition from A to B is
prevented by the excluded volume. Only the activity coefficient of the
interacting species, $\gamma_B$, is increased with respect to the
ideal behaviour. With respect to the previous exercise, the failure of
Eq.~\ref{eq:erroneous:check} is even more dramatic: $\vartheta$
decreases as a function of $N_T$, while the equation in question
predicts an increase:
\begin{equation}
  \frac{2 \gamma_A \gamma_B}{\gamma_A + \gamma_B}
  = \frac{N}{N-\frac{\bar{N}_B}{2}}
\end{equation}
The actual value of $\vartheta$ is lower than the ideal value, the
constant 1, while the prediction through Eq.~\ref{eq:erroneous:check}
is a deviation in the opposite direction from the ideal value.
\end{proof}

\begin{figure}
\includegraphics{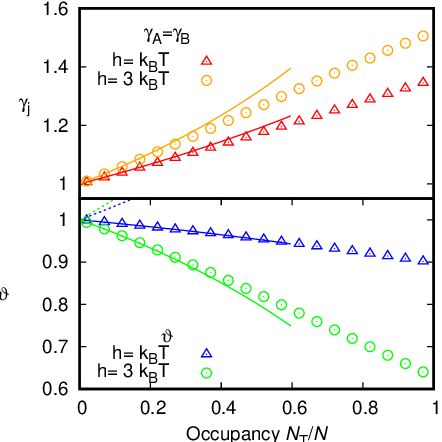}
  \caption{Monte Carlo results of
    Exercise~\ref{exercise:diff:repulsion}, on repulsive interaction
    between particles A and B. The solid curves are approximations,
    Eqs.~\ref{eq:ex:diff:repulsion:lambda} and
    \ref{eq:ex:diff:repulsion:Xi}.  The dashed curves are the values
    expected from the equation in question,
    Eq.~\ref{eq:erroneous:check}. }
  \label{fig:diff:repulsion}
\end{figure}
  
\begin{exercise}[Repulsive interaction between particles of different type]
  \label{exercise:diff:repulsion}
  Any number of particles is allowed to stay in each cell, however,
  particles A repel particles B in the same cell (and
  \emph{vice-versa}, thanks to the third law of mechanics). The
  Hamiltonian is thus:
  \begin{equation}
    H\left(s\right) = \sum_{j=1}^N h n_{A,j}\left(s\right) n_{B,j}\left(s\right)
  \end{equation}
  where $h$ is a positive interaction energy and $n_{A,j}(s)$
  (respectively, $n_{B,j}(s)$) is the number of particles A
  (respectively, B) in the $j$th cell in configuration $s$.  The
  solution can be obtained by Monte Carlo numerical calculation or by
  symbolic calculation, by approximating in the limit of small
  occupancy $N_T/N$.
\end{exercise}

\begin{proof}
  Fig.~\ref{fig:diff:repulsion} shows the results of a Monte Carlo
  numerical calculation. The approximated symbolic solution is
  obtained as follows.

  In the diluted limit, the large majority of the cells contain at
  most one particle. We can thus neglect cells with multiple particles
  and approximate:
  \begin{equation}
    \Gamma_A(s,j) = \delta_{n_{B,j}(s)\ge 1} e^{-h} + \delta_{n_{B,j}(s)=0} \\
  \end{equation}
  An analogous expression holds for $\Gamma_B$. Neglecting the effect
  of interactions on the particle distribution, we approximate:
  \begin{align}
    \left< \delta_{n_{B,j}(s)\ge 1} \right>_{N_T} & = \frac{\bar{N}_B}{N} \\
    \left< \delta_{n_{B,j}(s)=0} \right>_{N_T} & = \frac{N-\bar{N}_B}{N}
  \end{align}
  We thus find the quantities needed for the calculation of the
  activity coefficients $\gamma_j$:
  \begin{align}
    \left< \Gamma_A(s,j) \right>_{N_T} & = 1 - \frac{\bar{N}_B}{N}
    \left(1-e^{-h}\right) \\
    \left< \Gamma_B(s,j) \right>_{N_T} & = 1 - \frac{\bar{N}_A}{N}
    \left(1-e^{-h}\right)
  \end{align}
  For the calculation of $\vartheta$, it is useful to notice that:
  \begin{equation}
    \Gamma_A(s,j)^\alpha = \delta_{n_{B,j}(s)\ge 1} e^{-\alpha h} + \delta_{n_{B,j}(s)=0}
  \end{equation}
  and analogous expressions hold for other powers of $\Gamma_A(s,j)$,
  $\Gamma_B(s,j)$, and their products.

  By imposing the equilibrium through
  Eqs.~\ref{eq:def:mu:A:lambda:main} and
  \ref{eq:def:mu:B:lambda:main}, we find
  $\bar{N}_A=\bar{N}_B=N_T/2$. The results are:
  \begin{align}
    \gamma_A & = \gamma_B =
    \frac{1}{ 1 - \frac{N_T}{2N}\left(1-e^{-h}\right) }
    \label{eq:ex:diff:repulsion:lambda}
    \\
    \vartheta & = \frac
        { 1 - \frac{N_T}{2N} \left(2-e^{-\alpha h}-e^{-(1-\alpha) h}\right) }
        { 1 - \frac{N_T}{2N} \left(1-e^{-h}\right) }
    \label{eq:ex:diff:repulsion:Xi}
  \end{align}
  These approximated results are shown in
  Fig.~\ref{fig:diff:repulsion}: they approach the results of the
  Monte Carlo calculation in the limit of small occupancy $N_T/N$.

  The activity coefficients $\gamma_j$ are larger than 1 due to the
  repulsive nature of the interaction: the interaction of each
  particle participates with a positive additional term to the
  Hamiltonian and hence to the free energy.  Instead, $\vartheta$
  decreases: if two particles are present in a cell, they are more
  likely of the same type (both A or both B), and the transition of
  one of them is hindered, because it would lead to two different
  particles in the same cell (one A and one B), repelling each other
  and eventually increasing the mechanical energy.

  Like in the previous cases, we see that Eq.~\ref{eq:erroneous:check}
  is violated. Indeed, the predicted $\vartheta$ from the equation in question is:
  \begin{equation}
    \frac{2 \gamma_A \gamma_B}{\gamma_A + \gamma_B}
    = \frac{1}{ 1 - \frac{N_T}{2N}\left(1-e^{-h}\right) }
    \label{eq:repulsion:in:question}
  \end{equation}
  which increases as a function of $N_T$, while $\vartheta$ decreases.
\end{proof}

\begin{figure}
\includegraphics{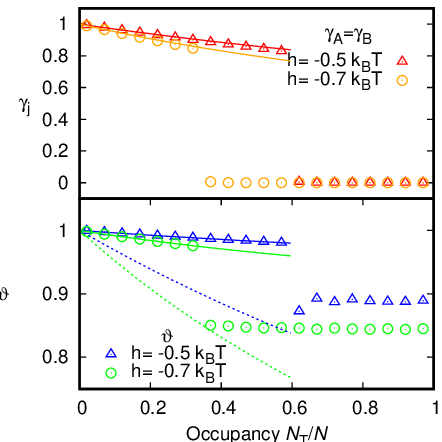}
  \caption{Monte Carlo results of
    Exercise~\ref{exercise:diff:attraction}, on attractive interaction
    between particles A and B. The solid curves are approximations,
    Eqs.~\ref{eq:ex:diff:repulsion:lambda} and
    \ref{eq:ex:diff:repulsion:Xi}.  The dashed curves are the values expected
    from the equation in question, Eq.~\ref{eq:erroneous:check}.}
  \label{fig:diff:attraction}
\end{figure}

\begin{exercise}[Attractive interaction between particles of different type]
  \label{exercise:diff:attraction}
  This exercise is analogous to
  Exercise~\ref{exercise:diff:repulsion}, but the interaction is an
  attraction, thus $h<0$.  The solution can be obtained by Monte Carlo
  numerical calculation or by symbolic calculation, by approximating
  in the limit of small occupancy $N_T/N$.
\end{exercise}

\begin{proof}
  The Monte Carlo calculation and
  Eqs.~\ref{eq:ex:diff:repulsion:lambda} and
  \ref{eq:ex:diff:repulsion:Xi} can be applied to this exercise, using
  $h<0$. The results are shown in Fig.~\ref{fig:diff:attraction}.

  The activity coefficients $\gamma_j$ are decreased by the
  interaction: every particle that interacts introduces a negative
  interaction term into the mechanical energy, in turn appearing in
  the free energy. The interaction also decreases $\vartheta$: cells
  with two different types of particles (one A and one B) are more
  likely than other combinations, and the transition of one of them is
  hindered, since it would lead to the loss of an attractive
  (negative) interaction energy.

  In Fig.~\ref{fig:diff:attraction}, it can be noticed that the
  approximations of Eqs.~\ref{eq:ex:diff:repulsion:lambda} and
  \ref{eq:ex:diff:repulsion:Xi} hold for small occupancy. However, the
  Monte Carlo results show an abrupt drop in both $\gamma_j$ and
  $\vartheta$. The reason is that, above a given threshold, it becomes
  very likely to have almost all the particles collapsing in the same
  cell. This is an unphysical phenomenon, however, it is similar to
  the phase transition that can actually take place in attractive
  systems.

  The value of $\vartheta$ predicted by Eq.~\ref{eq:erroneous:check} has the
  same expression of the previous exercise,
  Eq.~\ref{eq:repulsion:in:question}; in this case, since $h$ is
  negative, it predicts a decrease as a function of $N_T$. Actually,
  $\vartheta$ decreases, however, the equation in question appears to be
  quantitatively violated, as can be noticed in
  Fig.~\ref{fig:diff:attraction}.
\end{proof}

\section{Conclusion}

We propose a general scheme of a toy model, describing the equilibrium
of two types of reacting particles, possibly in the presence of
interactions. The interactions are defined by a Hamiltonian and the
thermodynamic quantities are evaluated based on statistical
mechanics. The exchange reaction rate is evaluated by modelling the
system in analogy with the collision theory used to justify the
empirical Arrhenius law. We provide general expressions for
calculating the activity coefficients and the exchange reaction rates
from the mathematical expression of the interactions.

The educational goal is to show that, even in a so simple system, it
is not possible to predict the exchange reaction rate from the
thermodynamic parameters, in particular, from the activity
coefficients. The students are invited to solve exercises, each
specifying a different interaction among particles. The exercises can
be solved by students having a minimal background in statistical
mechanics; knowledge of Monte Carlo numerical calculation can also be
exploited.

The results of the exercises show the absence of a general rule
connecting the activity coefficients with the exchange reaction rate.
In particular, the exercises prove that the substitution of activities
in the place of concentrations in mass-action equations is wrong,
although it may appear justified in specific cases, e.g. by the
transition state theory. The presented toy model can be used to
clarify that the activities cannot be thought of as ``more precise
concentrations''.

During the initial stages of chemistry courses, kinetics is usually
introduced for ideal systems, using mass action laws, based on
concentrations.  The general case (outside ideality) is later
explained by using thermodynamics, for chemical equilibrium; the
chemical potentials are introduced, clarifying that they can
deviate from ideality in the presence of interactions and that, in
some cases, the deviation can still be modelled, e.g. under excluded
volume interactions or under the Debye-H\"uckel theory.  The
activities and the activity coefficients are introduced as useful
mathematical shortcuts. We suggest that, at this stage, it is useful
to explicitly discuss the kinetics of non-ideal systems, clearly
explaining that there are methods to approach the problem, but all of
them require some microscopic model of the reaction and the
thermodynamic quantities are not enough to fully characterize the
systems. Care should be taken in order to avoid that the students jump
to unjustified extrapolations: the activities are not ``more precise
than concentrations'' in general, nor are the activity coefficients
``corrections to the concentrations'' in the context of kinetics.

From a broader point of view, the proposed toy model can be used to
show that thermodynamics does not predict rates. Under equilibrium,
the exchange rates are perfectly defined, and they are connected to
the behaviour of fluctuations by the fluctuation-dissipation
theorems~\cite{lebellac}. However, the knowledge of an independent
physical quantity, of kinetic nature, is always needed to calculate
the rates: thermodynamics does not have the concept of time.

\bibliographystyle{unsrt}
\bibliography{interactions}

\end{document}